\documentclass{ws-ijmpd}

\makeatletter
\def\wsdraftnote{}
\let\trimmarks\relax
\makeatother

\usepackage[super]{cite}
\usepackage{xcolor}
\usepackage[verbose,hypertexnames=false]{hyperref}
\hypersetup{colorlinks=false,allbordercolors=blue,pdfborderstyle={/S/U/W 1}}
\usepackage{url}

\usepackage{amsmath,amssymb,mathtools,bm}
\usepackage{mathrsfs}
\usepackage{physics}
\usepackage{comment}

\newcommand{\cM}{\mathcal M}
\newcommand{\cI}{\mathscr I}

\newcommand{\cQ}{\mathcal Q}

\newcommand{\Lie}{\mathcal L}
\newcommand{\bV}{\boldsymbol\upsilon}

\newcommand{\Iplus}{\mathscr I^+}

\begin{document}

\pagestyle{empty}

\markboth{Simon Pekar}
{Carrollian Conformal Dynamics and Flat-Space Holography}

%
\catchline{}{}{}{}{}
%

\title{Carrollian Conformal Dynamics and Flat-Space Holography}
\author{Simon Pekar}
\address{International School for Advanced Studies (SISSA),\\
Via Bonomea 265, 34136 Trieste, Italy.\\[5pt]
Istituto Nazionale di Fisica Nucleare (INFN), Sezione di Trieste,\\
Via Valerio 2, 34127 Trieste, Italy.\\[5pt]
\href{mailto:spekar@sissa.it}{spekar@sissa.it}}

\maketitle

\begin{abstract}
The conformal boundary of asymptotically flat spacetime is null and therefore carries a Carrollian rather than Lorentzian geometry. This changes the relation between symmetry, variational principles and evolution equations in an essential way: unlike a Lorentzian Levi--Civita connection, a torsion-free Carrollian connection compatible with the conformal structure contains genuine independent data. We develop a metric-affine formulation of conformal dynamics that keeps track of these connection degrees of freedom through hypermomenta. Local Carroll boosts, rotations, Weyl transformations and diffeomorphisms then imply a set of covariant evolution equations. We apply this framework to Einstein gravity in four-dimensional asymptotically flat spacetimes. Penrose's conformal compactification identifies null infinity as a conformal Carrollian manifold; the undetermined transverse symmetric part of its compatible connection is the asymptotic Bondi shear. After fixing a Bondi-van der Burg-Metzner-Sachs (BMS) frame and identifying the Carrollian momenta and hypermomenta with the Bondi data, the general Carrollian evolution equations reproduce the standard evolution equations for the mass and angular-momentum aspects. The result provides a boundary-intrinsic derivation of the gravitational evolution equations and a geometric realization of radiative degrees of freedom in flat-space holography.

This is a proceedings contribution to the ``Athens Workshop in Theoretical Physics: 10th Anniversary", held at the National and Kapodistrian University of Athens on December 17-19 2025.
\end{abstract}


\clearpage

\tableofcontents

\section{Introduction}

At the classical level, fluid/gravity \cite{Hubeny:2011hd} provides a useful example of how gravitational dynamics can be reformulated in terms of a boundary theory. It is a particular instance of the AdS/CFT correspondence between Anti-de Sitter spacetimes and conformal field theories, which reformulates the dynamics of a gravitational theory in terms of degrees of freedom of a Lorentzian fluid living on its codimension-one boundary. In asymptotically anti-de Sitter spacetimes, the boundary is timelike and its intrinsic geometry is Lorentzian; Einstein's equations can consequently be encoded in the covariant conservation and trace properties of a boundary stress tensor \cite{Balasubramanian:1999re,deHaro:2000vlm}. The asymptotically flat case is qualitatively different. Its conformal boundary is null, the induced metric is degenerate, and the appropriate boundary kinematics is Carrollian~\cite{Leblond,SenGupta:1966qer,Duval:2014uva,Ciambelli:2018wre,Ciambelli:2019lap}. At the same time, gravitational radiation reaches null infinity and carries energy and angular momentum away from the system. The asymptotic equations are therefore evolution equations rather than ordinary conservation laws.

For four-dimensional asymptotically flat gravity, the Bondi mass and angular-momentum aspects obey the evolution equations \cite{Bondi:1962px,Sachs:1962wk,Barnich:2010eb}
\begin{equation} \label{eq:intro-bms}
\begin{split}
\partial_u M &=\tfrac14 D_A D_B N^{AB}-\tfrac18 N_{AB} N^{AB}+\tfrac18 D^2 R, \\
\partial_u N_A &= \partial_A M + \tfrac14 C_{AB} \partial^B R-\tfrac14 N^{BC} D_A C_{BC}
+\tfrac1{16} \partial_A(C_{BC}N^{BC}) \\
&\qquad +\tfrac12 D^B\left(D_{[A} D^C C_{B]C}+C^C{}_{[A}N_{B]C}\right),
\end{split}
\end{equation}
where $u$ is the retarded time coordinate, giving rise to the null generators $\partial_u$ of future null infinity, $D_A$ is the covariant derivative on the cuts $u = \text{ctt.}$ of $\Iplus$ with topology $S^2$, $R$ their Ricci scalar, and $C_{AB}$ is the asymptotic shear. These equations encode the flux-balance laws for the asymptotic charges, among which the Bondi mass-loss formula describing the decrease of the total mass under emission of gravitational waves encoded in $N_{AB} = \partial_u C_{AB}$, the Bondi news.

The purpose of this proceeding is to give a geometric exploration of the boundary structure of four-dimensional asymptotically flat spacetimes and of the way in which they encode gravitational radiation. We then use this structure to derive the evolution equations Eqs.~\eqref{eq:intro-bms} intrinsically at null infinity. The central point is that, unlike in Lorentzian geometry, the compatible Carrollian connection is not uniquely determined by the metric data. Even after imposing torsionlessness and the appropriate compatibility conditions, a symmetric transverse part of the Carrollian connection remains arbitrary. At null infinity this apparent geometric ambiguity acquires a direct physical interpretation: it encodes the asymptotic shear. Once this independent connection data is included in the variational principle, its response is encoded in the hypermomenta as response to fluctuations of the connection \cite{Hehl:1976kj,Iosifidis:2020gth}. Diffeomorphism invariance subsequently yields covariant evolution equations whose BMS-frame expression reproduces the Einstein equations at null infinity.

\section{Lorentzian conformal dynamics}

Consider a relativistic field theory with fields $\phi^\alpha$ on $(d+1)$-dimensional Minkowski spacetime $\mathbb R^{d,1}$. From the Lagrangian $L[\phi^\alpha]$ one may define the canonical stress-tensor
\begin{equation}
    t^\mu{}_\nu:=\frac{\partial L}{\partial(\partial_\mu\phi^\alpha)}\partial_\nu\phi^\alpha-\delta^\mu{}_{\nu}L.
\end{equation}
Translation invariance gives $\partial_\mu t^\mu{}_\nu\approx0$, where $\approx$ denotes equality on the equations of motion. Lorentz invariance gives the corresponding angular-momentum Noether identity, and after the standard Belinfante improvement one obtains a symmetric tensor. If the theory exhibits in addition conformal invariance (dilations and special conformal transformations), then there exist further improvements that make the stress-tensor traceless \cite{Iosifidis:2025sjx}.

\subsection{Metric reformulation}

The previous derivation relies on Noether's first theorem and acquires a useful geometric reformulation. Suppose that the theory is minimally coupled to a Lorentzian background metric $g_{\mu\nu}$, so that the theory is now defined by an action $S[\phi^\alpha;g]$, integrated on the manifold $\mathcal M$. The variation can be written as
\begin{equation} \label{eq:metric-variation}
\delta S=\int_{\cM}\dd^{d+1}x\,\sqrt{-g}
\left(\mathcal E_\alpha\,\delta\phi^\alpha+\tfrac12 T^{\mu\nu}\delta g_{\mu\nu}\right),
\end{equation}
where $\mathcal E_\alpha$ represents the equations of motion and $T^{\mu\nu}$ is automatically symmetric. The metric transforms under infinitesimal diffeomorphisms as $\delta_\xi g_{\mu\nu} = 2\,\nabla_{(\mu} \xi_{\nu)}$, so that diffeomorphism invariance of the action $\delta_\xi S \approx 0$ (up to a boundary term) implies the covariant conservation equation
\begin{equation}
\nabla_\mu T^{\mu\nu} \approx 0,
\label{eq:lor-conservation}
\end{equation}
on-shell for the matter fields $\mathcal E_\alpha \approx 0$.
Thus, the covariant stress-tensor is the response to variations of the metric, and its conservation is associated with diffeomorphisms.

For a conformally invariant relativistic theory, one also has the Weyl symmetry $\delta_B g_{\mu\nu}=-2B g_{\mu\nu}$. If it is a variational symmetry, then on-shell
\begin{equation}
    g_{\mu\nu}T^{\mu\nu}\approx0,
\end{equation}
up to possible anomalies.\footnote{The latter are not present in three-dimensional Lorentzian CFTs, and we will assume that this property remains true in the case of three-dimensional Carrollian CFTs.} In flat space, this $T_{\mu\nu}$ becomes the improved traceless stress tensor familiar from conformal field theory.

The same logic will apply below once all background geometric sources are included in the variational principle: a boundary theory will inherit constraints and evolution equations from its local symmetries once all background geometric variables appearing in its variational principle have been correctly identified. In the rest of this section, we will rederive and generalize the previous formulas to the metric-affine case in the frame, as a preparation for the study of Carrollian manifolds.

\subsection{Frame-like metric-affine formulation}

Working at the level of a local coframe $\{\boldsymbol \theta^A\}$ on $T^*\mathcal M$, we can write the metric as $\boldsymbol g=\eta_{AB}\boldsymbol\theta^A\otimes\boldsymbol\theta^B$, where $\eta_{AB}$ is the $(d+1)$-dimensional flat Minkowski metric. The coframe carries the non-holonomy coefficients
\begin{equation}
    \dd \boldsymbol \theta^A + \tfrac12 c^A{}_{BC} \boldsymbol\theta^B \wedge \boldsymbol \theta^C = \boldsymbol 0.
\end{equation}
Local Lorentz transformations $\lambda_{[AB]}$, Weyl transformations $B$ and linearized diffeomorphism $\boldsymbol\xi$ act as
\begin{equation}
\delta_\lambda\boldsymbol\theta^A=\lambda^A{}_B\boldsymbol\theta^B,\quad \delta_B\boldsymbol\theta^A=-B\boldsymbol\theta^A,\quad \delta_\xi\boldsymbol\theta^A=\Lie_\xi\boldsymbol\theta^A.
\end{equation}

We introduce a spin-connection $\boldsymbol\nabla$ with coefficients $\boldsymbol\omega^A{}_B = \omega^A{}_{CB} \boldsymbol\theta^C$ in the local basis, and a Weyl-connection $\boldsymbol\alpha = \alpha_A \boldsymbol \theta^A$, transforming under local Lorentz and Weyl transformations
\begin{equation}
    \delta_\lambda \boldsymbol\omega^A{}_B = -\boldsymbol\nabla \lambda^A{}_B,\qquad \delta_B \boldsymbol\omega^A{}_B = \delta^A{}_B\,{\rm d}B, \qquad \delta \boldsymbol\alpha = -\dd B,
\end{equation}
while being a set of one-forms under diffeomorphisms. For a torsionless and Weyl-metric-compatible connection, \emph{i.e.}, imposing
\begin{equation}
    \boldsymbol \nabla \boldsymbol\theta^A = \dd \boldsymbol\theta^A + \boldsymbol\omega^A{}_B \wedge \boldsymbol\theta^B = \boldsymbol 0,\qquad \boldsymbol \nabla \boldsymbol g - 2 \boldsymbol\alpha \otimes \boldsymbol g = \boldsymbol 0,
\end{equation}
the frame formulation of the Levi--Civita theorem fixes the connection in terms of the metric and the Weyl one-form as
\begin{equation}
    \omega^A{}_{BC} = \tfrac12\left(c^A{}_{BC} + c_B{}^A{}_C + c_C{}^A{}_B\right) - \delta^A{}_B \alpha_C - \delta^A{}_C \alpha_B + \eta_{BC} \alpha^A.
\label{eq:spin-connection}
\end{equation}

One may be worried that the Weyl connection $\boldsymbol\alpha$ adds extra information compared to the usual metric formulation. It can realize that the latter is a purely auxiliary field under local special conformal transformations \cite{Fiorucci:2026nyg}
\begin{equation} \label{eq:special-conformal}
    \delta_Z \boldsymbol\alpha = Z_A \boldsymbol\theta^A,\qquad \delta_Z \boldsymbol \omega^A{}_B = Z^A \boldsymbol\theta_B - Z_B \boldsymbol\theta^A -\delta^A{}_B Z_C \boldsymbol\theta^C,
\end{equation}
such that one can always reach a gauge in which $\boldsymbol\alpha = \boldsymbol 0$ (the metric gauge).

\subsection{General evolution equations for metric-affine theories}

The variational principle \eqref{eq:metric-variation} becomes more instructive in metric-affine theories, where the connection is treated as an independent background field. On-shell for the dynamical matter fields, the general variation reads
\begin{equation}
    \delta S\approx\frac{1}{16\pi G}\int_{\cM}\boldsymbol\mu \left(\delta\boldsymbol\theta^A[\boldsymbol T_A]+\delta\boldsymbol\omega^A{}_B[\boldsymbol\Omega_A{}^B]\right),
\label{eq:lor-metric-affine-var}
\end{equation}
where $\boldsymbol\mu = \boldsymbol\theta^0 \wedge \dots \wedge \boldsymbol\theta^d$ is the volume form and the factor $(16\pi G)^{-1}$ is conventional. The vectors $\boldsymbol T_A = T^B{}_A \boldsymbol e_B$ and $\boldsymbol\Omega_A{}^B = \Omega_A{}^{CB} \boldsymbol e_C$ are respectively called the momenta conjugate to the coframe and the hypermomenta conjugate to the connection.

Local Lorentz and Weyl symmetries give algebraic identities relating the antisymmetric and trace parts of the momentum to covariant divergences of the hypermomentum
\begin{equation}
    T_{[AB]} = \boldsymbol{\mathcal D}[\boldsymbol\Omega_{[AB]}],\qquad T^A{}_A = - \boldsymbol{\mathcal D}[\boldsymbol\Omega_A{}^A],
    \label{eq:constraints}
\end{equation}
where $\boldsymbol{\mathcal D} = \boldsymbol\nabla + w \boldsymbol\alpha$ when acting on tensors of weight $w$. The last constraint can be interpreted as the requirement that no new degree of freedom is introduced in the direction of the Weyl-connection, as the latter is purely auxiliary.

Diffeomorphism invariance yields a modified conservation laws, in which the curvature couples to the hypermomentum
\begin{equation}
\boldsymbol{\mathcal D}[\boldsymbol T_A]=\boldsymbol{\mathcal R}^B{}_C[\boldsymbol e_A,\boldsymbol \Omega_B{}^C].
\label{eq:lor-metric-affine-eom}
\end{equation}
If the connection is Weyl-metric-compatible and torsionless, then the connection is not an independent quantity and takes the expression \eqref{eq:spin-connection}. The usual conservation law is recovered after substituting the Weyl--Levi--Civita expression and using the chain rule, and Eq.~\eqref{eq:lor-metric-affine-eom} can be recast as an exact conservation law for the improved stress-tensor $\boldsymbol{\tilde T}_A = \tilde T^B{}_A \boldsymbol e_B$ with
\begin{equation}
    \tilde T^B{}_A = T^B{}_A + \mathcal D_C \left(\Omega^{[B}{}_A{}^{C]} + \Omega^{[B|C|}{}_{A]} + \Omega_{[A}{}^{|B|C]} \right),
\end{equation}
which is symmetric and traceless by virtue of Eq.~\eqref{eq:constraints}, if and only if the identity $\Omega_A{}^{AB} + \Omega_A{}^{BA} - \Omega^B{}_A{}^A = 0$ is satisfied. The latter emerges naturally from requiring that the variational principle \eqref{eq:lor-metric-affine-var} is invariant under special conformal transformations displayed in Eq.~\eqref{eq:special-conformal}.

The takeaway is that conservation of an energy-momentum tensor is not a consequence of diffeomorphism invariance alone. It relies also on the critical assumption that the connection contains no independent physical source with in addition to the metric. If the connection has independent components, the corresponding response functions enter the evolution equations and the evolution equations acquire curvature-flux terms. This is precisely the mechanism that will occur in Carrollian conformal geometry: unlike in the Lorentzian case, independent connection components always appear even after imposing torsionlessness and Weyl-compatibility with the degenerate conformal structure.

\section{Carrollian conformal dynamics}

A Carrollian structure on a $(d+1)$-dimensional manifold $\mathcal C$ consists of a degenerate spatial metric $\boldsymbol g$ and a preferred vector field $\bV$ spanning its kernel,
\begin{equation}
\boldsymbol g(\bV,\cdot)=\boldsymbol 0.
\end{equation}
The conformal Carroll structure with homogeneous scaling identifies
\begin{equation}
(\boldsymbol g ,\bV)\sim(\mathscr B^{-2} \boldsymbol g,\mathscr B \bV).
\label{eq:carroll-conformal}
\end{equation}
Due to the metric being degenerate, one is led to introduce an auxiliary one-form $\boldsymbol\tau$ (the clock-form) which is dual to $\bV$: $\boldsymbol\tau[\bV] = 1$. The clock form is determined up to a part which is transverse to $\bV$.\footnote{Transverse forms $\boldsymbol X$ are such that $\boldsymbol X[\bV] = 0$. Conversely, given a clock form $\boldsymbol\tau$, transverse vectors $\boldsymbol V$ are such that $\boldsymbol\tau[\boldsymbol V] = 0$.} The last transformation is called a Carroll boost, and emerges in the $c \to 0$ limit of a local Lorentz boost. Therefore, we will prefer to work in the first-order (frame-like) formulation of Carrollian manifolds.

\subsection{Carroll frames and compatible connections}

The vector $\bV$ generates the distinguished Carrollian evolution direction. Locally, we choose a Carroll--Cartan frame $\{\boldsymbol e_A\} = \{\bV,\boldsymbol e_a\}$ with $a=1,\dots,d$, and work with its dual coframe $\{\boldsymbol\theta^A\} = \{\boldsymbol\tau,\boldsymbol \theta^a\}$. The degenerate Carroll metric can be defined as $\boldsymbol g=\delta_{ab}\boldsymbol\theta^a\otimes\boldsymbol\theta^b$.
The Carroll coframe carries non-holonomy defined by
\begin{equation}
    \dd \boldsymbol\tau + \varphi_a \boldsymbol\tau \wedge \boldsymbol\theta^a + \varpi_{ab} \boldsymbol\theta^a \wedge \boldsymbol\theta^b = \boldsymbol 0,\qquad \dd \boldsymbol\theta^a - c^a{}_b \boldsymbol\tau \wedge \boldsymbol\theta^b + \tfrac12 c^a{}_{bc} \boldsymbol\theta^b \wedge \boldsymbol\theta^c = \boldsymbol 0,
\end{equation}
where the quantity $\varphi_a$ carries the meaning of an acceleration, while $\varpi_{ab}$ measures the failure of the distribution generated by $\boldsymbol\tau$ to be integrable.
We decompose the symmetric projection of the quantity $c^a{}_b$ into a traceless part and a trace
\begin{equation}
    c_{(ab)} := \xi_{ab} + \tfrac1{d} \delta_{ab} \theta,
\end{equation}
where $\xi_{ab}$ is known as the geometric shear, \emph{i.e.} the traceless part of the extrinsic curvature of $\boldsymbol g$ in the direction of $\bV$, and $\theta$ is the expansion.

The local Carroll coframe transforms under infinitesimal rotations $r^{[ab]}$, boosts $\lambda_a$ and Weyl $B$ as follows
\begin{equation}
\delta \boldsymbol \tau = - B \boldsymbol\tau - \lambda_a \boldsymbol\theta^a, \qquad\delta \boldsymbol \theta^a= - B \boldsymbol \theta^a + r^a{}_b \boldsymbol \theta^b,
\end{equation}
so that the Carrollian metric $\boldsymbol g$ and vector field $\bV$ transform only under Weyl transformations, with weight $-2$ and $+1$ respectively.

Let now $\boldsymbol\omega^A{}_B$ be a Carrollian connection associated to the covariant derivative $\boldsymbol\nabla$ and impose that the torsion two-forms vanish
\begin{equation} \label{eq:vanishing-torsion}
\begin{split}
\boldsymbol\nabla \boldsymbol\tau &:= \dd\boldsymbol\tau+\boldsymbol\omega^0{}_0\wedge\boldsymbol\tau+\boldsymbol\omega^0{}_a\wedge\boldsymbol\theta^a = \boldsymbol 0,\\
\boldsymbol\nabla \boldsymbol\theta^a &:= \dd\boldsymbol\theta^a+\boldsymbol\omega^a{}_0\wedge\boldsymbol\tau+\boldsymbol\omega^a{}_b\wedge\boldsymbol\theta^b = \boldsymbol 0.
\end{split}
\end{equation}
A Weyl-compatible Carroll connection obeys in addition
\begin{equation}
\boldsymbol\nabla \boldsymbol g - 2\boldsymbol\alpha\otimes \boldsymbol g= \boldsymbol 0,
\qquad
\boldsymbol\nabla \bV+\boldsymbol\alpha\otimes \bV=\boldsymbol 0,
\label{eq:carroll-metricity}
\end{equation}
involving a Weyl one-form $\boldsymbol\alpha$ transforming as usual $\delta_B \boldsymbol\alpha = -\dd B$.

Unlike the Lorentzian case, a torsion-free Weyl-compatible Carroll connection exists only when the geometric shear vanishes and the temporal Weyl connection is identified with the expansion
\begin{equation}
    \xi_{ab} = 0, \qquad \alpha_0 = \tfrac1{d} \theta.
\end{equation}
Even when this is satisfied, the connection is not uniquely determined since
\begin{equation}
    \beta_{(ab)} := \boldsymbol\omega^0{}_{(a} [\boldsymbol e_{b)}]
\end{equation}
is left unspecified by the conditions Eq.~\eqref{eq:vanishing-torsion} and Eq.~\eqref{eq:carroll-metricity}. Equivalently, torsion-free compatible Carroll connections form an affine space modeled on transverse symmetric rank-two tensors. All in all, the connection reads
\begin{equation}
    \boldsymbol\omega^0{}_0 = - \boldsymbol \alpha,\quad \boldsymbol\omega^0{}_a = \beta_a \boldsymbol\tau + \beta_{ab} \boldsymbol\theta^b, \quad \boldsymbol\omega^a{}_0 = \boldsymbol 0,\quad \boldsymbol\omega^a{}_b = -c^a{}_b \boldsymbol\tau + \omega^a{}_{cb} \boldsymbol\theta^c,
\end{equation}
where $\beta_a = \varphi_a - \alpha_a$, $\beta_{ab} = \beta_{(ab)} - \varpi_{[ab]}$ and
\begin{equation} \label{eq:connection-transverse}
    \omega^a{}_{bc} = \tfrac12 \left(c^a{}_{bc} + c_b{}^a{}_c + c_c{}^a{}_b \right) - \delta^a{}_b \alpha_c - \delta^a{}_c \alpha_b + \delta_{bc}.
\end{equation}
It transforms under local gauge variations as
\begin{equation}
    \delta \boldsymbol \omega^a{}_b = - \boldsymbol\nabla r^a{}_b + \delta^a{}_b \dd B, \qquad \delta \boldsymbol\omega^0{}_a = \boldsymbol\nabla \lambda_a + \lambda_a \boldsymbol\omega^0{}_0 + r_a{}^b \boldsymbol\omega^0{}_b, \qquad \delta \boldsymbol\omega^0{}_0 = \dd B.
\end{equation}

As in the Lorentzian case, one may worry about the additional degrees of freedom encoded in $\beta_{ab}$ and $\alpha_a$. Using the Carrollian version of special conformal isometries, it can be seen that $\alpha_a$ (alternatively $\beta_a$) is shifted by spatial special conformal transformations and that $\beta^a{}_a$ is shifted by temporal special conformal transformations. Consequently, one may impose a gauge where $\alpha_a = 0$ (equivalently $\beta_a = 0$), and $\beta^a{}_a = 0$. The remaining degrees of freedom contained in $\beta_{\langle ab \rangle}$ constitute however genuine additional data \cite{Fiorucci:2026nyg,Hartong:2026bvh}. 

\subsection{Variational principle}

Let now $S[\phi^\alpha;\boldsymbol\theta,\boldsymbol\omega]$ be an effective action for unspecified Carrollian fields $\phi^\alpha$ coupled to the Carrollian conformal geometry. The frame and connection are treated as independent background fields, whereas $\phi^\alpha$ are dynamical. On-shell for matter fields,
\begin{equation}
\delta S\approx\frac{1}{16\pi G}
\int_{\mathcal C} \boldsymbol\mu\left(\delta\boldsymbol\theta^A[\boldsymbol T_A]+\delta\boldsymbol\omega^A{}_B[\boldsymbol\Omega_A{}^B]\right),
\label{eq:carroll-variation}
\end{equation}
with $\boldsymbol\mu=\boldsymbol\tau\wedge\boldsymbol\theta^1\wedge\cdots\wedge\boldsymbol\theta^d$.
The vectors $\boldsymbol T_A = T^B{}_A \boldsymbol e_B$ and $\boldsymbol\Omega_A{}^B = \Omega_A{}^{CB} \boldsymbol e_C$ are Carrollian momenta and hypermomenta, respectively, where $T^0{}_0$, $T^a{}_0$, $T^0{}_a$ and $T^b{}_a$ have the meanings of energy density, energy flux, momentum density and transverse stress-tensor respectively. The response to independent variations of the connection is the hypermomentum $\boldsymbol \Omega_A{}^B$. Its inclusion in the variational principle is not the result of a choice, but a requirement of Carroll geometry: since compatible Carroll connection are not uniquely fixed by $(\mathcal C,\boldsymbol g,\bV)$ the connection encodes genuine degrees of freedom.

In the context of a gravitational bulk dual for $d=2$, this free tensor is determined by extrinsic data. More precisely, using the induced connection obtained from the Weyl--Levi--Civita connection of the conformally compactified bulk, the tensor $\beta_{\langle ab\rangle}$ is the boundary value of the deviation tensor of an auxiliary outgoing null congruence, which is related to the asymptotic shear.

Local boost invariance gives
\begin{equation}
    T^a{}_0\approx-\boldsymbol{\mathcal D}[\boldsymbol\Omega_0{}^a],
\label{eq:boost}
\end{equation}
where $\boldsymbol{\mathcal D}=\boldsymbol\nabla+w\boldsymbol\alpha$ on a quantity of Weyl weight $w$. Thus the energy flux need not vanish: local Carroll boosts do not in general imply that the energy flux vanishes. Rather, it is fixed by the response to the independent connection encoded in hypermomentum. Local rotations and Weyl transformations give
\begin{equation}
T_{[ab]} \approx\boldsymbol{\mathcal D}[\boldsymbol\Omega_{[ab]}],\qquad T^A{}_A \approx-\boldsymbol{\mathcal D}[\boldsymbol\Omega_A{}^A].
\label{eq:rotation-weyl}
\end{equation}
These are the Carrollian analogues of the symmetry and trace identities familiar from relativistic conformal dynamics. Diffeomorphism invariance gives the central Carrollian evolution equations,
\begin{equation}
\boldsymbol{\mathcal D}[\boldsymbol T_0]
\approx \boldsymbol{\mathcal R}^A{}_B[\bV,\boldsymbol\Omega_A{}^B],
\qquad
\boldsymbol{\mathcal D}[\boldsymbol T_a]
\approx \boldsymbol{\mathcal R}^B{}_C[\boldsymbol e_a,\boldsymbol \Omega_B{}^C],
\label{eq:carroll-evolution}
\end{equation}
where $\boldsymbol{\mathcal R}^A{}_B$ is the curvature two-form of the Carrollian Weyl connection.

The right-hand sides are characteristic of the departure from ordinary conservation. If the hypermomentum associated with the independent part of the connection $\beta_{ab}$ does not vanish, these equations may not reduce to genuine covariant conservation laws. This is the Carrollian version of the general metric-affine identity discussed in the previous section, but with an important difference: the independent connection sector is forced on us by the kinematics of degenerate geometry itself.

The resulting interpretation is useful for holography. The momentum $\boldsymbol T_A$ carries the response associated with the charge aspects of the boundary theory, while $\boldsymbol \Omega_A{}^B$ carries the response associated with the radiative source. In the BMS frame, the evolution equations should therefore be understood as Carrollian conformal identities in the presence of an independently varying boundary connection.

\section{Application to gravitational evolution equations}

A four-dimensional asymptotically flat bulk metric in retarded coordinates $(r,u,x^A)$ may be written in the Bondi gauge \cite{Bondi:1962px,Sachs:1962wk,Barnich:2010eb} as
\begin{equation}
\dd s^2=-e^{2\beta}\frac Vr\dd u^2-2e^{2\beta}\dd u\,\dd r
+g_{AB}\,(\dd x^A-U^A\dd u)\,(\dd x^B-U^B\dd u),
\end{equation}
with
\begin{equation}
g_{AB}=r^2\,\gamma_{AB}+r\,C_{AB}+ \mathcal O(1),
\qquad
\gamma^{AB}C_{AB}=0,
\end{equation}
and the fall-offs $V = \mathcal O(r)$, $\beta = \mathcal O(r^{-2})$ and $U^A = \mathcal O(r^{-2})$. Einstein's equations at leading order force $\partial_u \gamma_{AB}$ to vanish, while the Bondi news reads simply $N_{AB}=\partial_u C_{AB}$. The spatial curvature of $\gamma_{AB}$, denoted by $R$, appears at leading order in $V$.

The metric contains also the Bondi mass aspect $M$ and angular-momentum aspect $N_A$ appearing at order $\mathcal O(1/r)$ in $\dd u^2$ and $\dd u\,\dd x^A$ respectively. Einstein's equations yield the BMS evolution equations, which take the form of Eq.~\eqref{eq:intro-bms}. Our goal is to recover these equations intrinsically, without importing the bulk Einstein equations in Bondi coordinates.

\subsection{Conformal compactification and Carroll geometry}

Consider an asymptotically flat four-dimensional spacetime with metric $\dd s^2$. Penrose's conformal compactification introduces an unphysical metric \cite{Penrose:1962ij}
\begin{equation}
    \dd s^2_\text{unphys}=\Omega^2 \dd s^2,
\qquad
\Omega\big|_{\cI}=0,
\qquad
\dd\Omega\big|_{\cI}\neq0,
\end{equation}
where $\Omega$ is a conformal factor which vanishes only at $\cI$. At null infinity, the normal is itself tangent. We denote the induced generator field by $\bV$. The pullback of the unphysical metric at $\cI$, denoted by $\boldsymbol g$ is degenerate along $\bV$:
\begin{equation}
\boldsymbol g(\bV,\cdot)=\boldsymbol 0.
\end{equation}
Moreover, because the conformal factor $\Omega$ is not fixed uniquely, but only its conformal class, the boundary data are defined up to
\begin{equation}
\boldsymbol g\mapsto \mathscr B^{-2} \boldsymbol g,
\qquad \bV \mapsto \mathscr B \bV.
\end{equation}
Therefore, $\cI$ naturally carries a conformal Carroll structure with $d=2$.

Einstein's equations at leading order force the trace-free part of the extrinsic curvature $\Lie_{\bV} \boldsymbol g$ to vanish, which in a Cartan frame is equivalent to $\xi_{ab} = 0$, so that a Carrollian connection which is both torsion-free and Weyl-metric-compatible exists. Finally, Ashtekar's analysis \cite{Ashtekar:1981hw} shows that the pullback of the unphysical Levi-Civita connection to the boundary fixes the remaining degrees of freedom in the Carrollian connection in terms of the Bondi shear, giving $\beta_{ab} = -\tfrac12 C_{ab}$.

\subsection{BMS frame-fixing and residual transformations}

To compare the intrinsic Carrollian equations with the BMS equations \eqref{eq:intro-bms}, we start by fixing the BMS frame
\begin{equation} \label{eq:BMS-frame}
\boldsymbol\tau=\dd u
\quad \Leftrightarrow\quad
\varphi_a=0=\varpi_{ab},
\qquad
\Lie_{\bV} \boldsymbol\theta^a = 0 \quad \Leftrightarrow\quad c_{ab}=0.
\end{equation}
The first two conditions make the cuts integrable and places restrictions on general Carroll-boost invariance. While $\xi_{ab} = 0$ already by virtue of Einstein's equations, further setting $\theta$ and $c_{[ab]}$ to zero fixes part of the local Weyl and rotation freedom to parameters invariant along $\bV$. We will also choose the gauge $\alpha_a = 0 = \beta^a{}_a$, attained thanks to spatial and temporal special conformal transformations (the latter is sometimes called subleading Weyl transformation \cite{Fiorucci:2025twa}). We note that in our choice of BMS frame \eqref{eq:BMS-frame}, the covariant derivative $\boldsymbol{\mathcal D}$ acting on transverse vector fields coincides with the covariant derivative $\nabla_a$ built upon $\boldsymbol\omega^a{}_b$ only.

A general boundary diffeomorphism $\boldsymbol\xi=f \bV +Y^a \boldsymbol e_a$ will take us away from the BMS frame, but one can accompany it by compensating local boost, rotation and Weyl transformations, so as to define the total variation
\begin{equation}
\delta_{\boldsymbol\xi}:=\Lie_{\boldsymbol\xi}+\delta_{\lambda(\boldsymbol\xi)}+\delta_{r(\boldsymbol\xi)}+\delta_{B(\boldsymbol\xi)}.
\end{equation}
We find that the transformations that preserve the frame satisfy
\begin{equation}
\bV[Y^a]=0,
\qquad
\bV[f]=\tfrac12 \nabla_a Y^a,
\qquad
\nabla_{\langle a}Y_{b\rangle}=0,
\label{eq:bms-parameters}
\end{equation}
and the local gauge parameters are fixed as
\begin{equation}
\lambda_a(\boldsymbol\xi)=\boldsymbol e_a[f],
\qquad
B(\boldsymbol\xi)= \bV[f],
\qquad
r_{ab}(\boldsymbol\xi)=\delta_{c[a}\boldsymbol\theta^c[\Lie_{\boldsymbol\xi}\boldsymbol e_{b]}].
\label{eq:compensators}
\end{equation}
Thus, $Y^a$ is a conformal Killing vector on each celestial cut, generating $\mathfrak{so}(3,1)$, while $f = T + \frac{u}2 \nabla_a Y^a$ generates the Abelian ideal of supertranslations $\mathscr C^\infty(S^2)$ with parameter $T$. Together, they generate the $\mathfrak{bms}_4$ algebra.

\subsection{The Bondi shear as boundary connection data}

In the BMS frame, one simply has
\begin{equation} \label{eq:boundary-connection}
\boldsymbol\omega^0{}_0=\boldsymbol 0,\qquad \boldsymbol\omega^0{}_a=-\tfrac12 C_{ab}\boldsymbol\theta^b,\qquad \boldsymbol\omega^a{}_0=\boldsymbol 0,\qquad \boldsymbol\omega^a{}_b=\omega^a{}_{cb}\boldsymbol \theta^c,
\end{equation}
where $C_{ab}$ is symmetric and traceless and $\omega^a{}_{bc}$ is given by Eq.~\eqref{eq:connection-transverse} for $\alpha_a = 0$. This gives a clear geometric interpretation of the Bondi shear: it is not introduced as an additional tensor on top of the Carroll geometry, but appears as the undetermined part of the compatible boundary connection. In other words, the two radiative gravitational degrees of freedom are encoded intrinsically in the connection itself.

The action of the residual symmetry on the shear follows directly from its appearance in the connection. One finds
\begin{equation}
    \delta_{\boldsymbol\xi} C_{ab}
=f\,\bV[C_{ab}]+(\Lie_Y C)_{ab}-\bV[f]\,C_{ab}
-2\nabla_{\langle a}\nabla_{b\rangle}f.
\label{eq:shear-transform}
\end{equation}
The inhomogeneous last term shows that the shear transforms as a connection component rather than as an ordinary tensor.

In the Bondi frame, the curvature of this connection has only a few non-zero components, which read
\begin{equation}
    \mathcal R^a{}_{bcd} = \delta^a{}_{[c} \delta_{d]b} R,\qquad \mathcal R^0{}_{a0b} = -\tfrac12 N_{ab}, \qquad \mathcal R^0{}_{abc} = - \nabla_{[b} C_{c]a},
\end{equation}
where $N_{ab} := \bV[C_{ab}]$ coincides with the Bondi prescription.

\subsection{Holographic dictionary and evolution equations}

We now specify the minimal holographic dictionary that makes the intrinsic Carrollian evolution equations \eqref{eq:carroll-evolution} in the BMS frame \eqref{eq:boundary-connection} identical to the BMS evolution equations \eqref{eq:intro-bms}.

As a preliminary remark, let us stress that --- in analogy with the Fefferman--Graham case --- the components of the stress-tensor, being of Weyl weights $+3$, are expected to encode Coulombic information about the spacetime at hand, like its mass or angular momentum aspects, collectively known in this context as the BMS momenta. On the other hand, the components of the hypermomentum tensor being objects of Weyl weight $+2$, they cannot be related in a local way to the BMS momenta. Instead, we will see that they encode information about radiation, as they are identified with components of the boundary Riemann tensor. Part of the boundary connection being free, the associated evolution equations incorporate a flux term that cannot be recast as conservation laws, which is a well-known fact. However, the rest of the boundary connection being related to the frame geometry, the split between momenta and hypermomenta is not unique, which mirrors the fact that the split of these equations between flux and charge is ambiguous \cite{Wald:1999wa}.

We now give the holographic dictionary that maps the Carrollian momenta and hypermomenta to the Bondi data and makes the intrinsic evolution equations reproduce the BMS evolution equations. The non-zero components of the hypermomenta are set as
\begin{equation}
\Omega_a{}^c{}_b := \mathcal R^{0c}{}_{ba} = \nabla_{[a} C_{b]}{}^c,\qquad
\Omega_0{}^{ab} := - 2 \mathcal R^{0a}{}_0{}^b + \mathcal R^{ca}{}_c{}^b = N^{ab}+\tfrac12 R\delta^{ab},
\label{eq:hyper-dictionary}
\end{equation}
which are expressed in terms of the non-vanishing components of the intrinsic Riemann tensor. The Carrollian momenta are
\begin{subequations}
\begin{align}
T^0{}_0&:=4M,\label{eq:Pi-dict}\\
T^0{}_a&:=2N_a+\tfrac1{16}\nabla_a(C^{bc}C_{bc}),\label{eq:P-dict}\\
T^a{}_0&:=- \nabla_b N^{ab}-\tfrac12 \nabla^a R,\label{eq:energy-flux-dict}\\
T_{ab}&:=\nabla^c \nabla_{[a}C_{b]c}
-\tfrac12N_{[a}{}^cC_{b]c}
-\tfrac14 R C_{ab}-2M\delta_{ab}.
\label{eq:stress-dict}
\end{align}
\end{subequations}
These expressions have to be interpreted as the flat-space analogue of the familiar relation between the holographic stress tensor and subleading coefficients in Fefferman--Graham expansions. Note that only $T^0{}_0$, $T^0{}_a$ and $T_{\langle ab\rangle}$ are freely specified. The other components of the Carrollian stress-tensor are fixed by invariance under local boosts, Weyl and rotation by Eqs.~\eqref{eq:boost}-\eqref{eq:rotation-weyl} and are identified with the geometry (frame and/or radiation).

The definition of the charges, defined as the boundary terms in the variation under diffeomorphisms, provides a further check. For a supertranslation/Lorentz pair $(f,Y^a)$, the boundary Noether charge on a cut $\Sigma\subset\Iplus$ is
\begin{equation}
\cQ_{(f,Y)}
=\frac{1}{16\pi G}\int_\Sigma \boldsymbol\mu_\Sigma
\left[4fM+2Y^aN_a+\tfrac1{16}Y^a\nabla_a(C_{bc}C^{bc})\right],
\label{eq:charges}
\end{equation}
where $\boldsymbol \mu_\Sigma = \tfrac12 \varepsilon_{ab} \boldsymbol\theta^a \wedge \boldsymbol\theta^b$, which coincides with the standard BMS charges in the corresponding frame \cite{Barnich:2010eb}. The radiative decorations in the dictionary \eqref{eq:P-dict} account for the difference between bare and covariant presentations of the asymptotic charges.

Finally, the on-shell variation \eqref{eq:carroll-variation} in the BMS frame contains the Ashtekar--Streubel term \cite{Ashtekar:1981bq}
\begin{equation}
    \vartheta_\text{AS} = \frac{1}{32 \pi G} \int_{\Iplus} \boldsymbol\mu\,\delta C_{ab} N^{ab},
\end{equation}
driving the non-conservation of the charges $\mathcal Q_{(f,Y)}$.

Note finally that, although $T_{\langle ab\rangle}$ is in principle free, it is not a completely new Coulombic data in our dictionary, as it is also fixed by geometry. In fact, one can deduce its expression by requiring the variational principle to be invariant under a \emph{new} type of symmetry corresponding a relaxed boundary gauge-fixing, leaving the transverse boundary geometry free to fluctuate, up to a fixed volume-form
\begin{equation}
    \delta_S \boldsymbol\theta^a = - S^a{}_b \boldsymbol\theta^b\quad \Rightarrow\quad \delta_S \left(\tfrac12 \varepsilon_{ab} \boldsymbol\theta^a \wedge\boldsymbol\theta^b\right) = \boldsymbol 0,
\end{equation}
with $S^{ab} = S^{\langle ab\rangle}$. This has two consequences. First, requiring that $\delta_S$ is a variational symmetry fixes the desired expression for the symmetric and traceless part of the transverse stress $T_{\langle ab \rangle} = -\tfrac14 R C_{ab}$. Second, taking $\delta_{S(\boldsymbol \xi)}$ into account, the residual transformations preserving the frame now allow for the full algebra of diffeomorphisms of the two-dimensional sphere, provided that $S_{\langle ab\rangle} = \nabla_{\langle a} Y_{b\rangle}$. In the asymptotic symmetries literature, these are known as the Campiglia--Laddha generalized BMS boundary conditions \cite{Campiglia:2014yka}, giving rise to superrotations. In a holographic setting, this would correspond to imposing Neumann boundary conditions on the transverse part of the metric, effectively fixing the corresponding part of the stress-tensor \cite{Compere:2008us,Compere:2019bua}. It would be interesting to come back to these issues in the context of holographic renormalization in asymptotically flat spacetimes.

\section{Discussion}

The present analysis suggests several directions for further work. A first question is whether the construction can be formulated in relaxed Bondi gauge \cite{Geiller:2022vto}. In general, the dictionary should be derived --- rather than postulated --- from a systematic holographic renormalization procedure at null infinity \cite{Hartong:2025jpp,Hartong:2026rbr}. Such a derivation would clarify the relation between the Carrollian momenta introduced here and the various proposed notions of Brown--York or holographic stress tensor for asymptotically flat spacetimes \cite{Kapec:2016jld,Chandrasekaran:2021hxc,Ruzziconi:2024kzo}. No bulk limiting procedure was needed in our construction: the derivation is intrinsic to the conformal Carroll boundary. However, a complementary approach would be to start with the holographic dictionary in AdS in a covariant Bondi gauge \cite{Compere:2019bua,Campoleoni:2023fug} (derived through holographic renormalization) and take the limit of vanishing cosmological constant. Unlike the Fefferman--Graham gauge, it allows to smoothly take the limit $\Lambda \to 0$, interpreted as a limit of the vanishing effective speed of light at the boundary. In asymptotically locally AdS$_4$ spacetimes, the change of gauge from Fefferman--Graham to Bondi should reflect the fact that, while still covariantly conserved, the original holographic stress-tensor receives geometric contributions under the form of components of the Cotton tensor, which ultimately give rise to flux terms in the flat limit.

Second, the particular form of the hypermomenta reveals what a microscopic Carrollian boundary theory must contain. A putative theory dual to four-dimensional gravity cannot couple only to the Carroll metric and observer field; it must also reproduce the response to the radiative connection sector. Constructing a microscopic theory with these specific properties would give a concrete realization of the source/radiative sector of flat-space holography. This geometric formulation should also have implications for scattering. Carrollian amplitudes naturally live at null infinity, and the present framework provides a covariant description of the background data to which the radiative sector couples. It remains to determine whether the connection–hypermomentum pair admits a direct interpretation in Carrollian or celestial amplitude bases \cite{Alday:2024yyj,Donnay:2022aba,Donnay:2022wvx,Nguyen:2023vfz,Isen:2026xoc}.

Finally, the distinction between conservation and evolution deserves further exploration. In Carrollian theories the failure of ordinary conservation is not necessarily a failure of symmetry: it can instead signal that the background contains independent connection data carrying flux. The present construction provides a concrete mechanism by which independent connection data modify the boundary Ward identities and produce evolution equations. More generally, the distinction between conservation and evolution deserves further study. In Carrollian theories, the failure of an ordinary conservation law need not signal a breaking of the underlying symmetry: it may instead reflect independent connection data carrying flux. A systematic understanding of the split between momenta and hypermomenta, and hence between charge and flux, remains an important open problem \cite{Donnay:2021wrk}.

\section*{Acknowledgements}

It is a pleasure to thank A.~Fiorucci, P.M.~Petropoulos and M.~Vilatte for the work this proceedings is based on, and also J.~Salzer for useful discussions. The author was supported by the \emph{Fonds Friedmann} run by the \emph{Fondation de l’\'Ecole polytechnique} until 31 October 2024. Since 1 November 2024, the author is supported by the \emph{European Research Council (ERC) Project 101076737 -- CeleBH}. Views and opinions expressed are however those of the author only and do not necessarily reflect those of the European Union or the European Research Council. Neither the European Union nor the granting authority can be held responsible for them. The author is also partially supported by \emph{INFN Iniziativa Specifica ST\&FI}.

\bibliographystyle{ws-ijmpd}
\bibliography{biblio.bib}

\end{document}